\documentclass[10pt,letterpaper]{article}
\usepackage{spconf,amsmath,graphicx,hyperref}
\usepackage[letterpaper,margin=1in]{geometry}
\usepackage{amssymb,bm}
\usepackage{booktabs}
\usepackage{orcidlink}
\usepackage{tabularx}
\usepackage{etoolbox}
\usepackage{cite}
\usepackage{microtype}
\hypersetup{hidelinks}
\patchcmd{\thebibliography}{\usecounter{enumi}}{\usecounter{enumi}\setlength{\itemsep}{0pt}\setlength{\parsep}{0pt}}{}{}

\newcommand{\Phat}{\widehat{P}}
\newcommand{\Shat}{\widehat{S}}

\newcommand{\CVaR}{\mathrm{CVaR}}

\title{PRIME-ANC: Path-Ratio-Informed Modeling for Efficient\\
Neural Filter Synthesis in Active Noise Control}

\name{\shortstack{Yaokun Huang\orcidlink{0000-0002-2031-1207}$^{1,2,*}$,
Chunyang Xu\orcidlink{0000-0002-2301-2173}$^{1,2,*}$,
Haowen Hua\orcidlink{0009-0005-4797-5715}$^{1,2,*}$,\\
Sen Lin\orcidlink{0009-0001-6115-1617}$^{1,2,*}$,
Shichao Hu\orcidlink{0009-0006-3809-5000}$^{1,2}$,
Mengyao Zhu\orcidlink{0000-0002-1069-4427}$^{1,2,\dagger}$}%
\thanks{$^*$These authors contributed equally and are listed in random order.}
\thanks{\textsuperscript{\textdagger}Corresponding author: \texttt{zhu.mengyao@\allowbreak suda.edu.cn}.}%
}

\address{
$^1$Center of Auditory and Language Intelligence,
Soochow University, Suzhou, China\\
$^2$School of Future Science and Engineering,
Soochow University, Suzhou, China
}

\begin{document}
\maketitle
\raggedbottom
\setlength{\parskip}{0pt}

\begin{abstract}
Changes in listener acoustics require active noise control (ANC) filters to be redesigned for new acoustic paths. We introduce PRIME-ANC, a shared neural synthesizer that learns a bounded, path-dependent log-magnitude correction to a regularized path-ratio base. Minimum-phase reconstruction and truncation produce finite-impulse-response (FIR) filters; training optimizes their noise-control performance. Across ten random training/test splits per dataset, PRIME-ANC achieves average held-out one-third-octave reductions of 18.81 and 17.76 dB over 50 Hz--5 kHz on a ten-path dataset and a public earphone database, respectively. On original earphone measurements, it improves upon the path-ratio base by 7.89 dB. Ablation studies support the contributions of both the analytic base and the path-dependent correction. Given calibrated paths for a held-out listener condition, PRIME-ANC generates a path-specific FIR without iterative optimization. With three Gauss--Newton updates, it reaches 21.43 dB reduction, approaching direct weighted least-squares design while producing lower amplification and root-mean-square control output.
\end{abstract}

\begin{keywords}
active noise control, acoustic-path conditioning,
amortized optimization, neural filter synthesis
\end{keywords}

\section{Introduction}

Listener anatomy and earphone fit alter acoustic paths, making rapid filter
redesign important in feedforward active noise control (ANC)
\cite{elliott1993anc,xiao2026robust}. Filtered-x least-mean-square (FxLMS)
and normalized FxLMS (FxNLMS) adapt filters online using a secondary-path
estimate \cite{kuo1996anc}. Alternatively, acoustic-model-based design
computes fixed filters for the target paths \cite{fabry2019fixed}.

For rapid responses to changing noise, selective fixed-filter ANC (SFANC)
selects predesigned filters \cite{shi2022sfanc}, while generative fixed-filter
ANC (GFANC) combines subfilters derived from a broadband filter
\cite{luo2023gfanc}. These filter resources remain system-dependent: GFANC
can reuse its trained network with subfilters matched to the new system
\cite{luo2025gfancImplementation}. End-to-end control-filter generation
(E2E-CFG) predicts complete filters from noise, but learns this mapping
under fixed paths and generally requires retraining for a new acoustic
environment \cite{yang2026e2ecfg}. Thus, transferring these noise-conditioned
methods to a new acoustic system still entails preparing suitable filter
resources or retraining the network.

Meta-learning reuses experience across acoustic conditions through learned
initial control filters \cite{shi2021mamlanc} and joint
control-filter/\allowbreak{}secondary-path initializations \cite{yang2026coinitialization}.
Regional initialization also improves attenuation at unseen source positions
before adaptation \cite{yang2025regional}. Primary and secondary paths determine
how disturbance and control signals combine at the error microphone.
We therefore combine cross-path learning with calibrated path information
in a shared generator trained on sparse support paths, directly synthesizing
a distinct causal finite-impulse-response (FIR) filter for each query
without retraining or query-specific optimization.

PRIME-ANC realizes this mapping by combining a regularized path-ratio
magnitude base with bounded, path-dependent log-magnitude corrections.
Nonminimum-phase secondary paths or insufficient primary-path delay can
prevent stable causal inversion \cite{zhu2019causal}. Training therefore evaluates
noise reduction, amplification and gain after minimum-phase reconstruction
and truncation; paired interpolation extends sparse support coverage.
The resulting FIRs also provide initializations for common Gauss--Newton
updates. Experiments on two datasets assess held-path synthesis, the
contributions of the base and conditioning, and the quality--design-cost
trade-off under refinement.

\section{PRIME-ANC Filter Synthesis}

\subsection{Acoustic paths and the filter-construction task}

Generator $G_{\bm\theta}$ is trained on a \emph{support set} $\mathcal S$
of path pairs. Each held-out \emph{query pair} $q\notin\mathcal S$ requires
a new filter. Let $p[n],s[n]$ denote primary and secondary impulse responses
and $w[n]$ a real, causal FIR. Reference noise $x[n]$ gives disturbance $d[n]$
and residual $e[n]$,
\begin{equation}
d=p*x,\qquad e=d-s*w*x ,
\label{eq:anc}
\end{equation}
where $*$ denotes convolution. Calibration measures residuals $e_0,e_g$ under probes $W_0=0$ and
$W_1=g$, a known nonzero gain. Hann-windowed transfer estimates
$\widehat H_{yx}$ divide pooled cross-spectra by regularized reference autospectra
and give
\cite{welch1967}
\begin{equation}
\Phat=\widehat H_{e_0x},\qquad
\Shat=(\widehat H_{e_0x}-\widehat H_{e_gx})/g .
\label{eq:pathid}
\end{equation}

Synthesis and evaluation are
\begin{equation}
\begin{aligned}
w_q&=G_{\bm\theta}(\Phat_q,\Shat_q),\\
e_q&=(p_q-s_q*w_q)*x .
\end{aligned}
\label{eq:construction}
\end{equation}
One network pass and deterministic realization produce $w_q$ without
optimization. Estimates guide synthesis and refinement; true support
paths supply supervision and true query paths supply residuals.

\begin{figure}[!t]
  \centering
  \includegraphics[width=0.98\columnwidth]{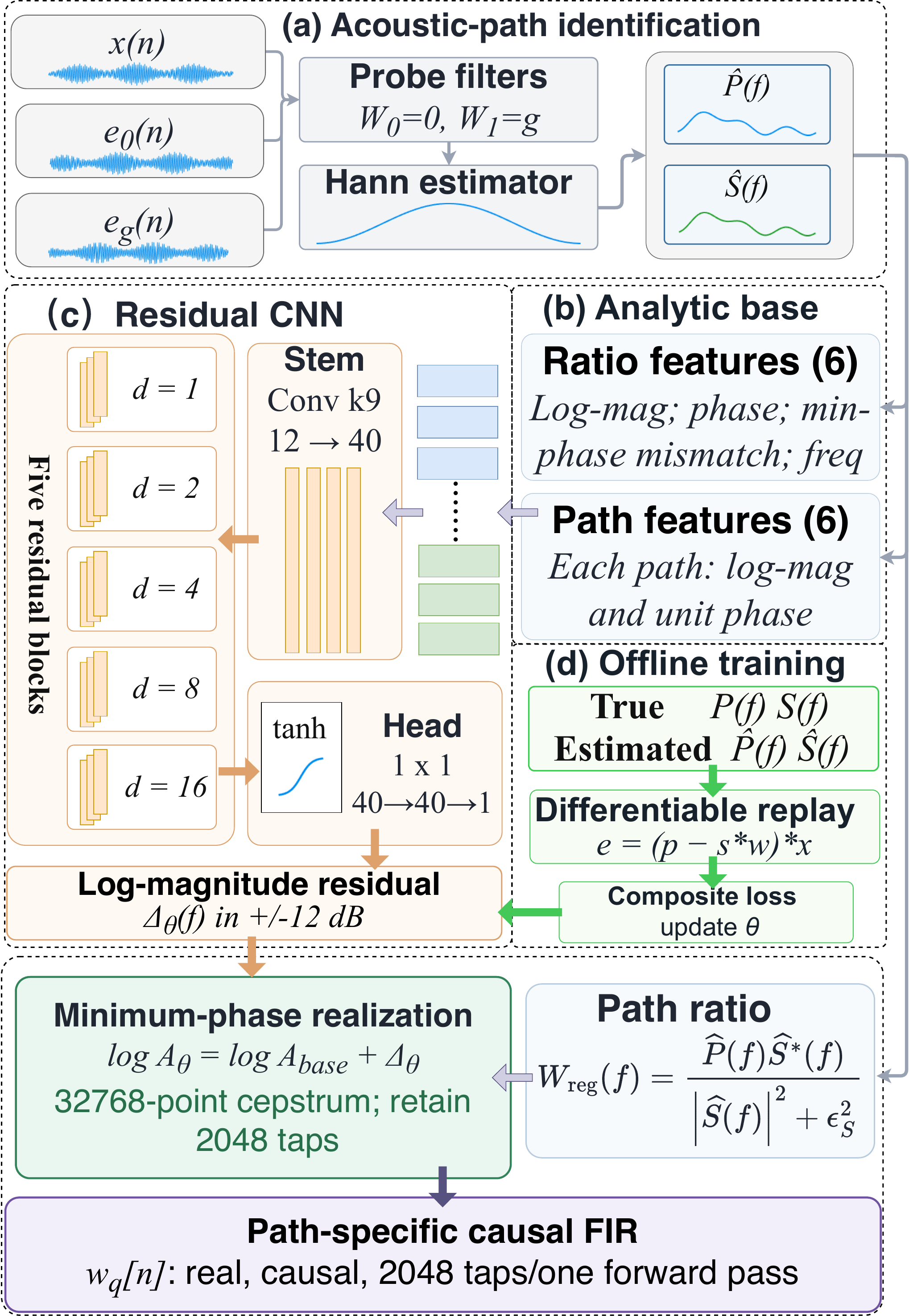}
  \caption{PRIME-ANC: path-conditioned FIR synthesis from the magnitude
  base $A_{\mathrm{base}}=|W_{\mathrm{reg}}|$ and a shared neural generator.}
  \label{fig:pipeline}
\end{figure}

\subsection{Path-ratio base and path-conditioned magnitude residual}

The ideal target response is $W_{\mathrm{ideal}}(f)=P(f)/S(f)$.
We regularize its calibrated estimate as
\begin{equation}
W_{\mathrm{reg}}(f)=
\frac{\Phat(f)\Shat^*(f)}{|\Shat(f)|^2+\epsilon_S^2},\qquad
\epsilon_S=10^{-4}\max_f|\Shat(f)|.
\label{eq:ratio}
\end{equation}
Regularization limits inversion near secondary-path spectral nulls
\cite{kirkeby1998deconvolution}. Its magnitude
$A_{\mathrm{base}}(f)=|W_{\mathrm{reg}}(f)|$ is the analytic base.
Realization with zero learned correction yields the
\emph{ratio-base FIR}.

A shared convolutional neural network (CNN) predicts $g_{\bm\theta}$
from calibrated path features, giving a bounded log-magnitude correction
(Fig.~\ref{fig:pipeline}):
\begin{equation}
\Delta_{\bm\theta}=\frac{12\ln 10}{20}\tanh g_{\bm\theta},\qquad
\log A_{\bm\theta}=\log A_{\mathrm{base}}+\Delta_{\bm\theta}.
\label{eq:residual}
\end{equation}
$\log$ is natural; corrections stay within $\pm12$ dB before realization.

Real-cepstral minimum-phase (MP) reconstruction supplies phase
\cite{oppenheim2010dsp}; an inverse Fourier transform retaining samples
0--2047 yields the real, causal FIR. Input phase guides magnitude
redistribution; reconstruction supplies output phase. Truncation can alter
magnitude and MP structure, so loss is evaluated on the realized FIR.

For conditioning, MP reconstruction of $A_{\mathrm{base}}$ supplies unit
phase $u_{\mathrm{MP}}$; $u=W_{\mathrm{reg}}/|W_{\mathrm{reg}}|$ is the ratio's
unit phase. Six ratio channels encode log magnitude, normalized frequency,
and real/imaginary parts of $u$ and the phase mismatch $u u_{\mathrm{MP}}^*$.
Six path channels encode log magnitudes and unit phases. Only log
magnitudes are standardized across frequency per path. The 151,401-parameter
CNN uses group normalization and sigmoid linear unit (SiLU) activations.

\subsection{Composite objective and support training}

Training uses 3-s noise clips and differentiable path convolution.
Let $D,E$ be the Hann-windowed short-time Fourier transforms (STFTs) of
disturbance and residual. Reduction in one-third-octave band
$b$ is $R_b=10\log_{10}[(\sum_{\mathcal I_b}|D|^2+\varepsilon)/
(\sum_{\mathcal I_b}|E|^2+\varepsilon)]$, where $\mathcal I_b$ contains its
STFT cells. Target-band noise reduction (NR) is
$|\mathcal B|^{-1}\sum_{b\in\mathcal B}R_b$ for band centers
$\mathcal B$ from 50 Hz to 5 kHz.
Negative $R_b$ indicates noise amplification (AMP), measured as $[-R_b]_+$.
We penalize its worst 10\% of bands from 50 Hz to 8 kHz using conditional
value-at-risk, $\CVaR_{0.1}$. The objective is
\begin{align}
\mathcal L={}&-\overline R_{50:5\mathrm{k}}
+0.25\,\CVaR_{0.1}([-R_b]_+) \notag\\
&+0.02\,\mathcal L_{24\mathrm{dB}}
+10^{-3}\overline{\Delta^2}+10^{-3}\mathcal L_{\mathrm{smooth}},
\label{eq:loss}
\end{align}
Here $\mathcal L_{24\mathrm{dB}}=\overline{[20\log_{10}|W|-24]_+^2}$
averages over the realized response's full Fourier grid;
$\overline{\Delta^2}$ penalizes log-magnitude correction, and
$\mathcal L_{\mathrm{smooth}}$ is the mean squared second difference
of realized taps. AdamW uses 6400 updates, batch
size 2, learning rate $5\times10^{-5}$, and weight decay $10^{-4}$;
the final checkpoint is evaluated.

To extend sparse training coverage, we optionally interpolate adjacent support
path pairs $i,j$ using one coefficient for both paths,
\begin{equation}
\widetilde p=(1-\alpha)p_i+\alpha p_j,\qquad
\widetilde s=(1-\alpha)s_i+\alpha s_j .
\label{eq:interp}
\end{equation}
Estimated inputs and true training paths share $\alpha$; held-out
conditions are unchanged.

\section{Experimental Setup}

\begin{figure*}[!t]
  \centering
  \includegraphics[width=0.96\textwidth]{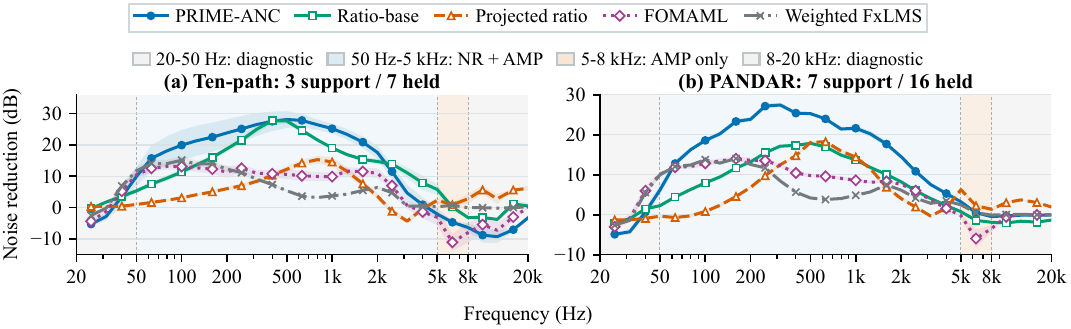}
  \caption{Frequency-resolved NR (mean $\pm1$ split SD).
  Gray: 40-s weighted FxLMS per path/noise (PANDAR: original measurements).}
  \label{fig:main_frequency}
\end{figure*}

\subsection{Datasets, training splits, and evaluation}

All experiments use 48 kHz and 2048-tap filters. Ten-path comprises ten
primary/secondary path pairs from Track 2 (ANC) of the 2026 CCF Advanced
Audio Technology Competition \cite{ccfaatc2026}. The public
PANDAR earphone acoustic-path database covers both ears of 23 participants
using the same earphone platform
\cite{liebich2019pandar}. We evaluate each measured secondary path and three
gain/delay and spectral/tail variants, with its primary path unchanged.
Both datasets use eight recordings from the Diverse Environments Multi-channel
Acoustic Noise Database (DEMAND) \cite{thiemann2013demand}, plus white and pink
noise, at root-mean-square (RMS) level 0.05.
Primary/secondary estimates have 2048/2048 taps on ten-path and 8192/2048 taps
on PANDAR. PANDAR calibration simulates probe responses from the available paths
and eight environmental recordings, with fixed paths across states and $g=0.35$. Hann transfer estimates
are pooled at six offsets. 
Supervised clips are drawn from 0--4 s; evaluation scores 7--10 s
after 0.5-s warm-up.

Ten random identity-based splits select
3/10 paths or 7/23 participants as support, holding out 7 paths or
16 participants. Ears and variants
stay with their participant. Each dataset/split has separately trained
networks; fitting and interpolation use support pairs only.
Each split adds 96 paired support mixtures, sampled with probability 0.75.
Ten-path interpolates adjacent supports ordered by their first principal
component; PANDAR interpolates adjacent support participants within each
ear and secondary-path condition. PRIME-ANC uses paired interpolation by default;
its PANDAR training averages 161.6 s per model.

Scores average six 0.5-s blocks: NR over 50 Hz--5 kHz and AMP
over 50 Hz--8 kHz, clipping $[-R_b]_+$ before averaging. Thus
5--8 kHz contributes to AMP only. Output RMS measures $w*x$.
We average noises, ear/variant conditions and three seeds within
each held path or participant, then held units within each split.
Means and sample standard deviations (SDs) describe ten paired,
equally weighted split means; the overlapping splits are not independent trials.

\subsection{Direct synthesis and iterative references}

Two analytic FIRs provide direct-construction baselines: ratio-base uses
PRIME-ANC's realization with zero correction; projected ratio retains
inverse-transform samples 0--2047 of complex $W_{\mathrm{reg}}$, without shifting. Both use
8192/32768-point Fourier grids on ten-path/PANDAR.
First-order model-agnostic
meta-learning (FOMAML) \cite{finn2017maml,shi2021mamlanc} learns one initial
FIR from true support paths, using 800 pooled Adam and 400 meta updates.
Table~\ref{tab:direct} reports the same pre-adaptation FIR for all queries,
without inner-loop updates.

Weighted minimum mean-square error (WMMSE) design supplies shared and
per-query filters by minimizing
\begin{equation}
\min_{w\in\mathbb R^{2048}}
\sum_{a\in\mathcal A}\sum_k
\omega_{a,k}|\Phat_{a,k}-\Shat_{a,k}\bm f_k^{\mathsf T}w|^2
+\lambda\|w\|_2^2 .
\label{eq:wmmse}
\end{equation}
Here $\bm f_k^{\mathsf T}$ is a 2048-tap Fourier row on a 32768-point
grid. Within each target band, reference-power weights are divided by
the band's Fourier-bin count, then normalized globally to unit mean
over active bins.
$\mathcal A=\mathcal S$ fits the shared filter; $\mathcal A=\{q\}$
fits query $q$, with $\lambda=10^{-4}\sum_{a,k}\omega_{a,k}|\Shat_{a,k}|^2$.
Shared filters use a direct Toeplitz solve. PANDAR query filters use
a double-precision CPU Toeplitz solve or single-precision GPU
preconditioned conjugate gradients (PCG) with fast Fourier transform
(FFT) products. PCG uses zero initialization, diagonal preconditioning,
relative residual tolerance $10^{-6}$ and budgets of 32--2048 iterations.

We adapt E2E-CFG's convolutional--Transformer backbone and direct
filter-generation approach \cite{yang2026e2ecfg} into a
path-conditioned direct-tap baseline. Eight calibrated-path channels replace
noise frames as inputs for predicting 2048 FIR taps.
The head receives means and SDs of $\log|\Phat|$, $\log|\Shat|$ and
$\log|W_{\mathrm{reg}}|$ to retain amplitude scale.
Training matches PRIME-ANC's physical loss weights, 6400 updates, batch
size 2, seeds and support-only paired interpolation on both datasets,
using AdamW at $5\times10^{-4}$ without the log-magnitude correction penalty.
The final checkpoint is evaluated, and we do not interpret this result
as a direct reproduction comparison with the original E2E-CFG setting.

Iterative FIR applies 8192 Adam steps
($5\times10^{-4}$) per query from
the ratio base to the reference-spectrum version of Eq.~\eqref{eq:loss},
complementing WMMSE's quadratic criterion (Table~\ref{tab:tradeoff}).
Single-precision Gauss--Newton (GN) refines PRIME, both analytic FIRs and
FOMAML, plus direct WMMSE on PANDAR. With generator parameters fixed,
it uses query estimates to minimize a tap-refinement objective
$\mathcal J$ combining mean log third-octave residual and smooth
worst-band risk (weight 0.05, temperature 1) on a
32768-point grid. Damping is $0.002\max(\bar h,10^{-6})+10^{-6}$,
where $\bar h$ is the mean unregularized normal diagonal. PCG allows
16 iterations with relative residual tolerance $10^{-6}$. Response-step
RMS is capped at $10^{1/20}-1$ times the query ratio-base RMS over active
bins. Backtracking over $1,1/2,1/4$ stops GN if no decrease reaches
$10^{-5}\max(|\mathcal J|,1)$. All starts share tap-update budget $K$;
$K=0$ is unrefined. Updates use FFT operators, fused PCG and CUDA
Graph replay.
FOMAML, WMMSE, iterative FIR and GN share a reference power spectrum
estimated from the ten noises over 0--4 s.

FxLMS and FxNLMS use true secondary paths and 50 Hz--5 kHz gradient
weighting, with the common scoring protocol. Figure~\ref{fig:main_frequency} shows weighted FxLMS on both datasets
(PANDAR: original measurements). For Table~\ref{tab:tradeoff}, each path's
filter is adapted on white noise and frozen for evaluation on ten noises.
Both share a validation-based stopping rule with a 240-s adaptation cap.

\begin{figure*}[!t]
  \centering
  \includegraphics[width=0.92\textwidth]{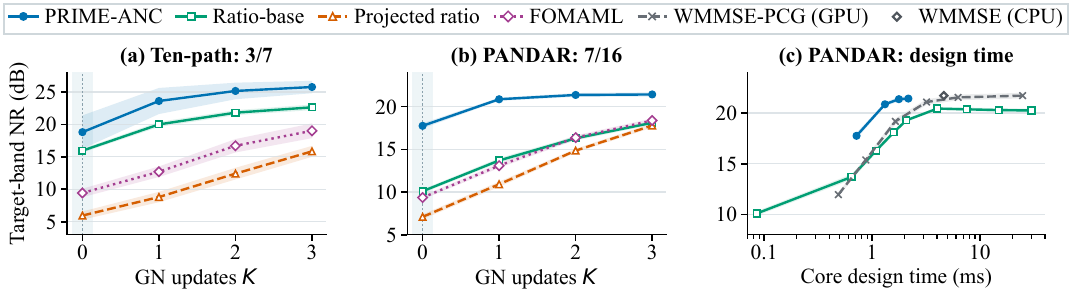}
  \caption{Common GN updates (a,b) and PANDAR core design time (c). Shading: $\pm1$ split SD.}
  \label{fig:initializers}
\end{figure*}

After runtime initialization, PANDAR single-query calls are timed three
times on an RTX 5090, except for direct WMMSE on one Threadripper PRO
7975WX CPU thread. Condition medians are averaged within participants,
seeds and splits as for NR. Core time covers query-specific filter
construction and optimization from resident features and Fourier paths,
including CPU-to-GPU transfer for WMMSE plus GN. Calibration, file access,
model loading and runtime initialization are excluded.

\section{Results and Discussion}

\subsection{Direct synthesis on held-out acoustic conditions}

Direct synthesis gains 2.857/7.663 dB NR over the ratio base on
ten-path/PANDAR (Table~\ref{tab:direct}), with positive
seed-averaged gains in 9/10 and 10/10 splits. Gains span much of
the lower and middle target bands (Fig.~\ref{fig:main_frequency});
mean AMP is 0.770/0.134 dB.
On original PANDAR measurements, PRIME improves every split, reaching
18.576 dB NR versus 10.686 dB for the ratio base, and 16.832 dB without interpolation.
Modified conditions gain 7.588 dB with interpolation.
Paired interpolation improves every split, adding 2.083/1.539 dB on
ten-path/PANDAR (Table~\ref{tab:direct}). Ten-path support/held ratios of
2/8--9/1 yield 16.28--22.76 dB NR; 3/7 uses sparse support.

\begin{table}[!t]
\caption{No query updates: NR (dB), mean $\pm$ SD over ten splits.}
\label{tab:direct}
\centering
\normalsize
\setlength{\tabcolsep}{3pt}
\begin{tabularx}{\columnwidth}{@{}>{\raggedright\arraybackslash}Xrr@{}}
\toprule
Method & Ten-path 3/7 & PANDAR 7/16\\
\midrule
PRIME-ANC, no interpolation & $16.72\!\pm\!2.08$ & $16.22\!\pm\!0.66$\\
PRIME-ANC, paired interpolation & $\mathbf{18.81}\!\pm\!2.57$ & $\mathbf{17.76}\!\pm\!0.30$\\
\addlinespace[1pt]
Ratio-base FIR (zero residual) & $15.95\!\pm\!0.29$ & $10.09\!\pm\!0.22$\\
Projected complex-ratio FIR & $5.95\!\pm\!0.63$ & $7.09\!\pm\!0.28$\\
FOMAML, pre-adaptation & $9.43\!\pm\!0.65$ & $9.35\!\pm\!0.22$\\
\addlinespace[1pt]
Support-shared WMMSE & $7.07\!\pm\!2.46$ & $8.81\!\pm\!0.27$\\
E2E-CFG (path-adapted)$^\dagger$ & $5.55\!\pm\!4.43$ & $9.75\!\pm\!0.42$\\
\bottomrule
\end{tabularx}
\par\vspace{2pt}
\begin{minipage}{\columnwidth}
\footnotesize\raggedright
$^\dagger$Path-conditioned adaptation of E2E-CFG; see Section~3.2.
\end{minipage}
\end{table}

\subsection{What the ratio base and conditioning contribute}

To isolate the base connection, four output controls
(Table~\ref{tab:mechanisms}) append summary statistics of
$\log|\Phat|$, $\log|\Shat|$ and $\log A_{\mathrm{base}}$ after the CNN trunk.
They use $\log A=b\log A_{\mathrm{base}}+(12\ln10/20)g_{\bm\theta}$,
with $b=1$ or $0$ and unbounded $g_{\bm\theta}$. Output phase is MP or
MP plus $\pi\tanh v_{\bm\theta}$ before the same FIR realization
(151,641/151,682 parameters). Within each phase construction, only the base connection changes;
features, initialization, augmentation, updates and seeds are matched.
All four replace the correction penalty with
$\overline{(\log|W|-\log A_{\mathrm{base}})^2}$; standard PRIME keeps
the bound and penalty in Eqs.~\eqref{eq:residual}--\eqref{eq:loss}.
We retain bounded PRIME across datasets, as the unbounded variant
improves PANDAR but degrades ten-path NR.

\begin{table}[!ht]
\caption{PANDAR construction and conditioning: NR/AMP (dB).}
\label{tab:mechanisms}
\centering
\normalsize
\setlength{\tabcolsep}{4pt}
\begin{tabular}{@{}lrr@{}}
\toprule
Filter construction & NR & AMP\\
\midrule
PRIME-ANC (standard) & 17.755 & 0.134\\
\addlinespace[1pt]
\multicolumn{3}{@{}l}{\emph{Output construction controls}}\\
Ratio-base skip, MP & 18.577 & 0.093\\
Ratio-base skip, MP + phase & 18.616 & 0.093\\
No skip, MP & 14.646 & 0.093\\
No skip, MP + phase & 14.655 & 0.093\\
\addlinespace[1pt]
\multicolumn{3}{@{}l}{\emph{Standard-model conditioning}}\\
Shuffled input phase & 12.593 & 0.289\\
Path-independent correction & 12.509 & 0.205\\
\bottomrule
\end{tabular}
\end{table}

The base adds 3.93/3.96 dB with MP/phase-corrected outputs in all
ten splits at similar AMP; phase correction adds $<0.04$ dB.
Path-independent correction and phase shuffling lose 5.246/5.162 dB
NR, supporting path-dependent correction and input phase use.

\subsection{Refinement and computational efficiency}

\begin{table}[!t]
\caption{PANDAR quality and core design time (NR/AMP in dB).
WMMSE + GN starts from the CPU direct solution.}
\label{tab:tradeoff}
\centering
\normalsize
\setlength{\tabcolsep}{2.8pt}
\begin{tabularx}{\columnwidth}{@{}>{\raggedright\arraybackslash}Xrrrr@{}}
\toprule
Method & NR & AMP & RMS & Time (ms)\\
\midrule
PRIME, direct & 17.755 & 0.134 & 0.302 & 0.717\\
PRIME + GN(1) & 20.869 & 0.136 & 0.303 & 1.317\\
PRIME + GN(2) & 21.370 & 0.149 & 0.306 & 1.772\\
PRIME + GN(3) & 21.430 & 0.156 & 0.307 & 2.169\\
Ratio-base + GN(4) & 19.286 & 0.412 & 0.310 & 2.079\\
\midrule
WMMSE, CPU direct & 21.722 & 0.351 & 0.502 & 4.623\\
WMMSE + GN(3) & 23.264 & 0.357 & 0.506 & 6.120\\
Iterative FIR & 21.992 & 0.165 & 0.449 & 12430.125\\
\bottomrule
\end{tabularx}
\par\vspace{3pt}
\begin{tabular*}{\columnwidth}{@{}l@{\extracolsep{\fill}}rrr@{}}
\toprule
Adaptive reference & NR & AMP & RMS\\
\midrule
\multicolumn{4}{@{}l}{\textit{40--120 s white-noise adaptation}}\\
FxLMS  & 19.497 & 0.125 & 0.150\\
FxNLMS  & 19.833 & 0.148 & 0.152\\
\bottomrule
\end{tabular*}
\end{table}

Common GN tests initialization under identical query estimates and
objective. PRIME leads analytic starts and FOMAML at all tested
budgets ($K=0$--3) on both datasets (Fig.~\ref{fig:initializers}a,b).
At comparable core times of about 2.1 ms, three GN updates from PRIME
achieve 2.14 dB higher NR than four GN updates from the ratio base
(Fig.~\ref{fig:initializers}c).
PRIME + GN(3) approaches direct WMMSE in NR with lower core time, AMP
and control RMS (Table~\ref{tab:tradeoff}). WMMSE + GN raises NR at higher cost.

\section{Conclusion}

PRIME-ANC combines an analytic path-ratio base with learned
path-dependent log-magnitude corrections to synthesize causal FIRs
from calibrated paths. Experiments on two datasets show gains over
the ratio-base FIR on held-out paths and listener conditions, while ablations support
both the base and path conditioning. On PANDAR, three GN updates
approach direct WMMSE in NR with lower measured core design time,
amplification, and control RMS.
These results highlight PRIME-ANC as a promising dynamic framework for earphone ANC, enabling rapid, listener-specific filter generation with lower design overhead and robust noise control performance.

\section{Acknowledgement}

This work was supported in part by the Science and Technology Program of Jiangsu Province (BZ2024062).

\bibliographystyle{IEEEbib}
\bibliography{references}

\end{document}